\documentclass[prb,amssymb,showpacs,twocolumn]{revtex4}

\usepackage{graphicx}

\def\avg#1{\left\langle#1\right\rangle}

\def\be{\begin{equation}}       \def\ee{\end{equation}}
\def\bea{\begin{eqnarray}}      \def\eea{\end{eqnarray}}
\def\ba{\begin{array} }
\def\ea{\end{array} }
\def\bnum{\begin{enumerate} }
\def\enum{\end{enumerate}}

\def\=>{\Rightarrow}
\def\>{\rightarrow}

\begin{document}

\title{Topological quantization of the spin Hall effect in two-dimensional
paramagnetic semiconductors}

\author{Xiao-Liang Qi$^1$, Yong-Shi Wu$^{2,1}$ and Shou-Cheng Zhang$^{3,1}$}

\affiliation{$^1$ Center for Advanced Study, Tsinghua University,
Beijing, 100084, China}
\affiliation{$^2$ Department of Physics, University of Utah, Salt
Lake City, UT 84112-0830}
\affiliation{$^3$ Department of Physics, McCullough Building,
Stanford University, Stanford, CA 94305-4045}

\begin{abstract}

We propose models of two dimensional paramagnetic semiconductors
where the intrinsic spin Hall effect is exactly quantized in
integer units of a topological charge. The model describes a
topological insulator in the bulk, and a ``holographic metal" at
the edge, where the number of extended edge states crossing the
Fermi level is dictated by (exactly equal to) the bulk topological
charge. We also demonstrate the spin Hall effect explicitly in
terms of the spin accumulation caused by the adiabatic flux
insertion.

\end{abstract}


\maketitle

\section{Introduction}

The intrinsic spin Hall effect is a novel phenomenon in condensed
matter physics, where a dissipationless spin current is proposed
to be induced by an external electric field. The effect has been
theoretically predicted both in $p$ doped semiconductors with
Luttinger type of spin-orbit (SO) coupling\cite{murakami2003} and
in $n$ doped semiconductors with Rashba type of SO
coupling\cite{sinova2004}. After these initial proposals, the
issue of the stability of the intrinsic spin Hall effect has been
intensely debated theoretically. It is now broadly believed that
the vertex corrections due to impurity scattering exactly cancels
the intrinsic spin Hall effect in the $n$ type Rashba
model\cite{inoue2004,raimondi2005,mishchenko2004}, while the
vertex corrections vanishes for the $p$ type
Luttinger\cite{murakami2004} and Rashba models\cite{bernevig2005}.
In the latter case, the impurity scattering does not affect the
intrinsic spin Hall effect in the clean limit. This conclusion is
also supported by extensive numerical calculations\cite{chen2005}.
Remarkably, the spin Hall effect has been experimentally
observed\cite{kato2004,wunderlich2005}. The experiment of Ref.
\cite{wunderlich2005} was carried out in a two dimensional hole
gas (2DHG) in the clean limit, and the effect is likely of the
intrinsic nature.

The next logical question in the study of the emerging field of
the spin Hall effect concerns the dissipationless nature of the
transport and possible quantization of the spin Hall conductivity.
Murakami, Nagaosa and Zhang proposed that the intrinsic spin Hall
effect can even exist in insulators where the Fermi level lies
within a band gap\cite{murakami2004B}. In a spin Hall insulator,
there is no charge current but spin currents, and the transport
can be completely dissipationless. Bernevig and Zhang have
proposed that the spin Hall effect can be quantized in two
dimensions\cite{bernevig2005A}. In their proposal, the Landau
levels arise from the gradient of the strain, rather than the
magnetic field.

In the present paper, we propose a new realization of the quantum
spin Hall effect (QSHE) by specializing the spin Hall insulator
model of Ref. \cite{murakami2004B} to two dimensions. In the
presence of mirror symmetry with respect to the $xy$ plane, the
system is shown to be a topological insulator characterized by a
momentum space winding number $n\in\mathbb{Z}$, with spin Hall
transport carried by gapless edge states in a cylindrical geometry,
in a way similar to the quantum (charge) Hall system. The evolution
of the edge states owing to the adiabatic flux insertion can be
traced by following Laughlin-Halperin argument for the integer
quantum Hall effect (IQHE)
\cite{laughlin1981,halperin1982,sheng2005}, and it can be related
explicitly to the spin accumulation at the boundary. Our model
therefore describes a bulk topological insulator, and a
``holographic metal" at the boundary, where the edge transport
properties precisely encode the bulk topological invariant.

The rest of this paper is organized as follows. In Sec. II we
introduce a systematic description of the quantum anomalous Hall
effect (QAHE) in the most general two-band model in two dimensions
that realizes the charge QHE without an external magnetic field. It
also provides a helpful mathematical preparation for understanding
the quantum spin Hall effect (QSHE). In Sec. III we show how the
QSHE emerges in a 2-dimensional spin Hall insulator with an
``inverted" band structure. Finally, Sec. IV is devoted to
conclusions and discussions.

\section{Quantized Anomalous Hall Effect}

To understand the topological quantization of the spin Hall
effect, we shall first introduce a general class of 2d models,
called quantum anomalous Hall insulators, in which the charge Hall
effect is topologically quantized in the absence of an external
magnetic field.

Historically, the first example of the QAHE was introduced by
Haldane\cite{Haldane1988}, which is a tight-binding model defined
on a honeycomb lattice with next nearest neighbor hopping and
stagger flux. (Recently, Kane and Mele generalized Haldane's model
and discussed the QSHE\cite{kane2004}.) Similar to the usual
quantum Hall effect, the QAHE is also a consequence of momentum
space topology\cite{thouless1982}, and is robust against local
perturbations. We will show the topological nature of the spin
Hall conductance in our general two-band model explicitly by a
Kubo formula calculation.


The most general two-band Hamiltonian describing a 2d
non-interacting system can be expressed in the following form:

\begin{eqnarray}
H= \sum_{\bf k} H({\bf k}),\quad H({\bf k})&=&\epsilon({\bf
k})+Vd_\alpha({\bf k})\sigma^\alpha , \label{QAHEHamiltonian}
\end{eqnarray}
where $\sigma^\alpha (\alpha=1,2,3)$ are the three Pauli matrices
and ${\bf k}=(k_x,k_y)$ stands for the Bloch wavevector of the
electron. The two bands may stand for different physical degrees
of freedom depending on the context. If they are the components of
a spin-1/2 electron, $d_\alpha({\bf k})$ describe the spin-orbit
coupling. If they correspond to the orbital degrees of freedoms,
then $d_\alpha({\bf k})$ describe the hybridization between bands.
The discussion below is completely independent of the physical
interpretation of the Hamiltonian (\ref{QAHEHamiltonian}), and
leads to a general understanding for the conditions for the QAHE.

The Hamiltonian (\ref{QAHEHamiltonian}) can be easily diagonalized
to obtain the two-band energy spectrum as $E_\pm({\bf
k})=\epsilon({\bf k})\pm Vd({\bf k})$, in which $d({\bf k})$ is the
norm of the 3-vector $d_\alpha({\bf k})$. The Hall conductivity can
be calculated using the standard Kubo formula to be
\begin{eqnarray}
\sigma_{xy}&=&\lim_{\omega\rightarrow
0}\frac i\omega Q_{xy}(\omega+i\delta)\ , \nonumber\\
Q_{xy}(i\nu_m)&=&\frac{1}{\Omega\beta}\sum_{{\bf k},n}{\rm
tr}\left(J_{x}({\bf k})G({\bf
k},i(\omega_n+\nu_m))\right. \nonumber\\
& &\left.\cdot J_y({\bf k})G({\bf k},i\omega_n)\right)\ ,
\label{kubo}
\end{eqnarray}
with the current operator ($i,j=x,y$)
\begin{eqnarray}
J_i({\bf k})&=&\frac{\partial H({\bf k})}{\partial
k_i}=\frac{\partial \epsilon({\bf k})}{\partial
k_i}+V\frac{\partial d_\alpha({\bf k})}{\partial
k_i}\sigma^\alpha,\nonumber\\
\label{currentoperator}
\end{eqnarray}
and $G({\bf k},i\omega_n)$ the Matsubara Green function.

From eqs. (\ref{kubo}) and (\ref{currentoperator}), the Hall
conductivity can be calculated straightforwardly. The detail of this
calculation is given in Appendix A, with the resulting $\sigma_{xy}$
given by
\begin{eqnarray}
\sigma_{xy}=\frac 1{2\Omega}\sum_{\bf k}\frac{\partial
\hat{d}_\alpha({\bf k})}{\partial k_x}\frac{\partial
\hat{d}_\beta({\bf k})}{\partial
k_y}\hat{d}_\gamma\epsilon^{\alpha\beta\gamma}
\left(n_+-n_-\right)({\bf k})\ ,
\label{conductcalculation}.
\end{eqnarray}
where $\hat{d}_\alpha({\bf k})=d_\alpha({\bf k})/d({\bf k})$ is
the unit vector along the direction of $d_\alpha({\bf k})$.
$\hat{d}_\alpha({\bf k})$ is singular if $d({\bf
k})=\sqrt{d_\alpha({\bf k})d^\alpha({\bf k})}$ vanishes for some
${\bf k}$. However, here and below we are always interested in the
insulating models, in which a full gap opens between the two bands
$E_+({\bf k})$ and $E_-({\bf k})$, thus $E_+({\bf k})-E_-({\bf
k})=2Vd({\bf k})>0$ for all ${\bf k}$. The gap opening condition
is written explicitly as
\begin{eqnarray}
\min_{{\bf k}\in {\rm BZ}}E_+({\bf k})>\max_{{\bf k}\in{\rm
BZ}}E_-({\bf k})\ .
\label{gapcondition}
\end{eqnarray}

\begin{figure}[tbp]
\begin{center}
\includegraphics[width=3.5in] {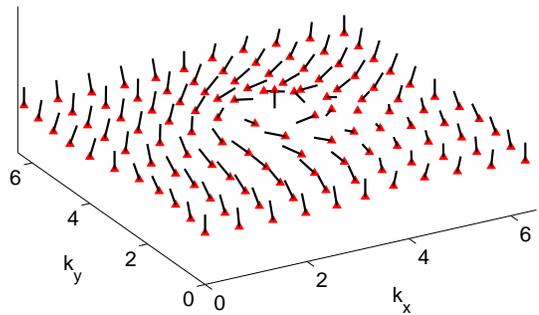}
\end{center}
\caption{The Skyrmion configuration of ${\bf\hat{d}}({\bf k})$ in
the Brilloune zone of the system (\ref{QAHEHReal}) with
$c=1,e_s=3.7,n=1$. The vector ${\bf \hat{d}}({\bf k})$ starts from
the north pole at the center of Brilloun zone and end at the south
pole at the zone boundary after covering the unit sphere once.}
\label{skyrmion}
\end{figure}

In this case, the system becomes a bulk insulator when the
chemical potential lies inside the gap, which implies $n_-({\bf
k})\equiv 1,\text{ } n_+({\bf k})\equiv 0$ for all ${\bf k}$ at
zero temperature. Under such condition and taking the
thermodynamic limit, the Hall conductivity
(\ref{conductcalculation}) can be simplified to
\begin{eqnarray}
\sigma_{xy}&=&-\frac{1}{8\pi^2}\int\int_{\rm FBZ} dk_xdk_y{\bf
\hat{d}}\cdot{\bf
\partial_x\hat{d}\times\partial_y\hat{d}}\label{conductQAHE}
\end{eqnarray}
which is a topological invariant defined on the first Brillioun
zone (FBZ), independent of the details of the band structure
parameters.\cite{yakovenko1990} Considering ${\hat{\bf d}({\bf
k})}: T^2\rightarrow S^2$ as a mapping from the Brillioun zone to
the unit sphere, the integrand ${\bf \hat{d}}\cdot{\bf
\partial_x\hat{d}\times\partial_y\hat{d}}$ is simply the Jacobian
of this mapping. Thus the integration over it gives the total area
of the image of the Brillioun zone on $S^2$, which is a
topological winding number with quantized value $4\pi n,n\in
\mathbb{Z}$. Thus the conductivity $\sigma_{xy}$ is always
quantized as $\sigma_{xy}=-n/2\pi$ when the mapping covers $S^2$
$n$ times. A schematic picture of a typical ${\bf \hat{d}}({\bf
k})$ configuration with winding number $n=1$ is shown in Fig.
\ref{skyrmion}. Although the single-electron states in this system
are very different from the Landau levels in the usual integer
quantum Hall effect (IQHE), the quantization of conductivity in
these two systems share the same topological origin, which can be
understood as Berry's phase in ${\bf k}$-space. The exact formula
(\ref{conductQAHE}) plays a key role for the QAHE which is similar
to that of the TKNN formula in the Landau level
problem\cite{thouless1982,kohmoto1985}. Consequently, both of them
are robust against weak disorder due to the topological
reason\cite{Niu1985}, which is well known for the IQHE case. The
general relationship between the momentum space topology and the
quantization of physical responses has been discussed extensively
by Volovik in Ref.\cite{Volovik2003}.

For an explicit discussion on the QAHE and the characteristics of
edge states, we consider the following choice of $d_a({\bf k})$ as
an example:
\begin{eqnarray}
d_x&=&\sin k_y,\qquad d_y=-\sin k_x,\nonumber\\
d_z&=&c\left(2-\cos k_x-\cos k_y-e_s\right)\label{QAHEHReal}
\end{eqnarray}
%

When $V/t$ in Hamiltonian (\ref{QAHEHamiltonian}) is large enough,
the insulator condition (\ref{gapcondition}) is satisfied and the
Hall conductivity can be shown to be

\begin{eqnarray}
\sigma_{xy}=\left\{\begin{array}{c c}1/2\pi, & 0<e_s<2\\-1/2\pi,
&2<e_s<4\\0,& e_s>4\text{ or }e_s<0\end{array}\right .
\end{eqnarray}
where the parameter $c$ is taken to be positive. Physically, this
model can be understood as a tight-binding model describing some
magnetic semiconductor with Rashba type SO coupling, spin dependent
effective mass and a uniform magnetization on $z$ direction. The
experimental realization of such QAHE will be discussed in future
works.

\begin{figure}[tbp]
\begin{center}
\includegraphics[width=2.5in] {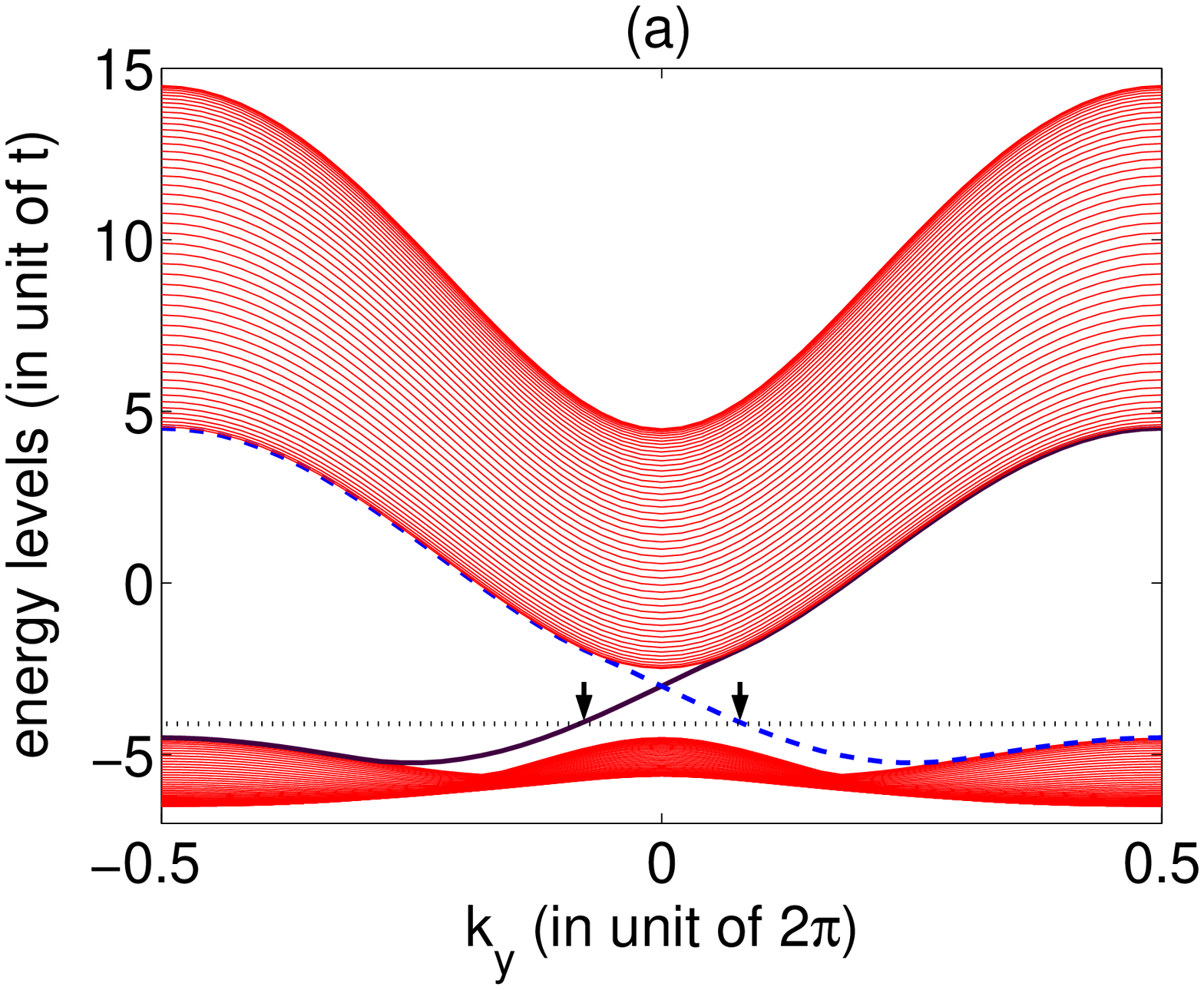}
\end{center}
\begin{center}
\includegraphics[width=2.5in] {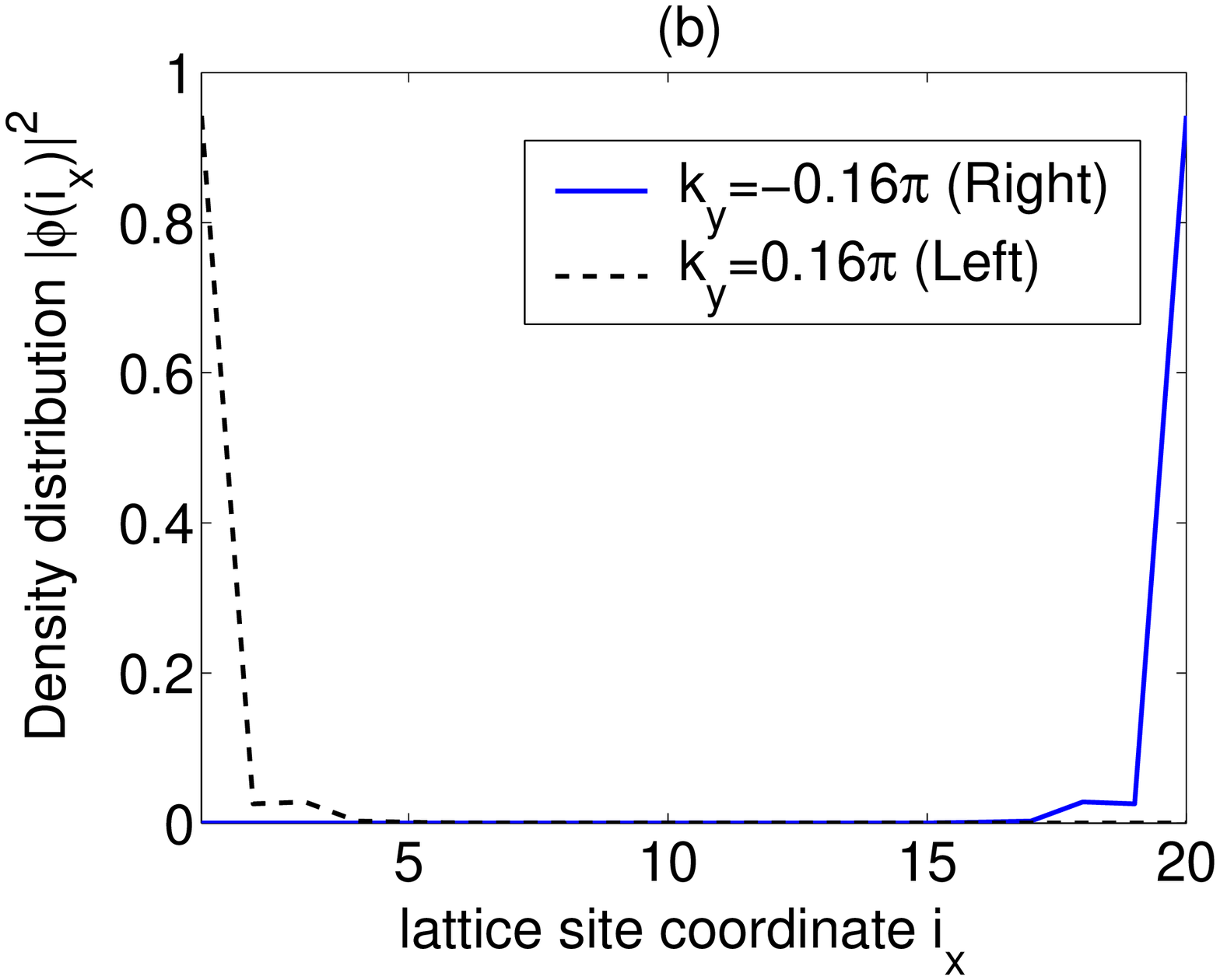}
\end{center}
\caption{(a) Energy spectrum of the QAHE system (\ref{QAHEHReal})
when the open boundary condition is imposed in $x$-direction. The
parameters are $c=1, t/V=1/3, e_s=0.5$. The solid and dashed lines
between two energy bands stand for the edge states at the right
and left edge, respectively. The horizontal dotted line shows a
typical chemical potential inside the gap. The two arrows mark the
two gapless edge excitation with momentum $k_y=\pm 0.16\pi$. (b)
the density distribution of the two edge states at fermi surface,
calculated for a $50\times 50$ lattice.} \label{QAHEexample}
\end{figure}

To show the behavior of edge states, one can define such a
Hamiltonian on a strip with the periodic boundary condition in
$y$-direction and open boundary condition in $x$-direction, with
vanishing wave function at $x=0,L+1$. In this case the Bloch
wavevector $k_y$ is a good quantum number and the single-particle
energy is a function $E_m(k_y), (m=1,\cdots,2L)$. A typical energy
spectrum is shown in Fig. \ref{QAHEexample}. For a given $k_y$,
there are $2L$ states, two of these are localized, while the rest
are extended. When the Fermi level lies in the bulk energy gap (as
represented in Fig. \ref{QAHEexample} by a horizontal dot line), the
only gapless excitations are edge states. Similar to the usual IQHE
case, the edge states have a definite chirality. In the present
case, the state localized at the left edge moves with velocity
$v_y<0$ and that at the right edge with $v_y>0$. This can be seen
directly from the dispersion relation of edge states. More
generally, when $\sigma_{xy}=n/2\pi$, there are $\left|n\right|$
chiral edge states on each edge.

\section{Quantum Spin Hall Effect}

Besides the interests in its own, the above discussions on the
QAHE also serve as a natural introduction to the QSHE. In fact, as
we shall see, the QSHE can be understood as two copies of the
QAHE; each breaks time reversal symmetry while the whole system
remains invariant under time reversal. In particular, let us
consider the light-hole and heavy-hole bands in a semiconductor
which can be described by the Luttinger model\cite{Luttinger1956}:
\begin{eqnarray}
H({\bf k})&=&\frac{\gamma_1+\frac{5}2\gamma_2}{2m}{\bf
k}^2+\frac{\gamma_2}{m}\left({\bf S\cdot k}\right)^2 \ ,
\end{eqnarray}
where the three components of ${\bf S}$ stand for the $4\times 4$
angular momentum matrices of a $J=3/2$ electron. According to
Ref.\cite{murakami2004A}, such a $4\times 4$ Hamiltonian can be
reexpressed by introducing five Dirac $\Gamma$ matrices
$\Gamma^a=\xi^{ij}_a\left\{S^i,S^j\right\},a=1,2,..,5$ as:
\begin{eqnarray}
H({\bf k})&=&\epsilon({\bf k})+Vd_a({\bf
k})\Gamma^a \ ,
\label{LuttingerGen}\\
\text{with }\epsilon({\bf k})&=&\frac{\gamma_1}{2m}{\bf
k}^2,\qquad V=\frac{\gamma_2}{m} , \nonumber\\
d_1&=&-\sqrt{3}k_yk_z,\qquad  d_2=-\sqrt{3}k_xk_z,\nonumber\\
d_3&=&-\sqrt{3}k_xk_y,\qquad
d_4=-\frac{\sqrt{3}}2\left(k_x^2-k_y^2\right),\nonumber\\
d_5&=&-\frac12\left(2k_z^2-k_x^2-k_y^2\right) ; \nonumber
\end{eqnarray}
and the matrices $\Gamma^a$ form the $SO(5)$ Clifford algebra
$\left\{\Gamma^a,\Gamma^b\right\}=2\delta^{ab}$. (More details
about the $SO(5)$ representation of Luttinger model can be found
in Appendix A of Ref.\cite{murakami2004A}.) By using the Clifford
algebra, the Hamiltonian (\ref{LuttingerGen}) can be diagonized to
obtain the doubly degenerate eigenvalues:
\begin{eqnarray}
E_{\pm}({\bf k})=\epsilon({\bf k})\pm V\sqrt{d_ad^a({\bf
k})}=\frac{\gamma_1\pm 2\gamma_2}{2m}{\bf k}^2 \ .
\end{eqnarray}

In the present work, we will focus on the spin Hall insulators
described by $\gamma_2>\gamma_1/2$,\cite{murakami2004B} and
specialize it to two dimensions. When the semiconductor described
by the Hamiltonian (\ref{LuttingerGen}) is made into a quantum
well in $z$-direction, the effective Hamiltonian can be obtained
by adding a potential well term $U(z)$ to the original Luttinger
Hamiltonian. When $U(z)$ is narrow enough, the system can be
considered as quasi-2D in the low energy sector, for which one can
write down a two-dimensional effective Hamiltonian $H_{\rm 2d}$.
The simplest way to obtain $H_{\rm 2d}$ is by replacing
$k_z,k_z^2$ in the original Hamiltonian (\ref{LuttingerGen}) by
their average in the lowest subband $\avg{k_z},\avg{k_z^2}$,
respectively. When the potential is symmetric, $U(z)=U(-z)$, the
parity symmetry with respect to x-y plane is respected and thus
$\avg{k_z}=0$, which means $d_1=d_2\equiv 0$, and the 2d
Hamiltonian can be simplified to
\begin{eqnarray}
H_{\rm 2d}&=&\epsilon({\bf k})+Vd_\alpha({\bf
k})\Gamma^{\alpha}, (\alpha=3,4,5) \label{H2d}\\
d_3&=&-\sqrt{3}k_xk_y,\qquad
d_4=-\frac{\sqrt{3}}2\left(k_x^2-k_y^2\right),\nonumber\\
d_5&=&-\frac12\left(2e_s-k_x^2-k_y^2\right).\nonumber
\end{eqnarray}
with $e_s\equiv\avg{k_z^2}$ and ${\bf k}=(k_x,k_y)$.

Noticing that $\Gamma_\alpha (\alpha=3,4,5)$ form a reducible
representation of an $SO(3)$ Clifford sub-algebra, it is natural
to see the similarity of the Hamiltonian (\ref{LuttingerGen}) to
the two component one (\ref{QAHEHamiltonian}) proposed in the
previous section. To see such a similarity explicitly, define
\[
\Gamma^{ab}=\left[\Gamma^a,\Gamma^b\right]/2i,\qquad (a,b=1,2,
\cdots,5)
\]
then
\begin{eqnarray}
\left[\Gamma^{12},\Gamma^\alpha\right]=0 (\alpha=3,4,5)
\Rightarrow \left[\Gamma^{12},H_{\rm 2d}\right]=0
\end{eqnarray}
Therefore, $\Gamma^{12}$ serves as a ``conserved spin quantum
number" even in the presence of the SO coupling. The eigenvalues
of $\Gamma^{12}$ are $\pm 1$ both with double degeneracy, and
$\Gamma^{12}$ nad all of $\Gamma^\alpha$ $(\alpha=3,4,5)$ can be
block-diagonalized simultaneously. Since they form a
representation of $SO(3)$ Clifford algebra, the new expression of
$\Gamma^a$ in the diagonal representation of $\Gamma^{12}$ can
always be chosen as
\begin{eqnarray}
\Gamma^{12}=\left(\begin{array}{c c} I & \\ &
-I\end{array}\right),\qquad\Gamma^{\alpha}=\left(\begin{array}{c c}
\sigma^{\alpha-2} &
\\ & -\sigma^{\alpha-2}\end{array}\right)
\end{eqnarray}
with $\alpha=3,4,5$, and $\sigma^\mu (\mu=1,2,3)$ are the Pauli
matrices. In this new representation, the 2D Luttinger Hamiltonian
(\ref{LuttingerGen}) is also block diagonal:
\begin{eqnarray}
H({\bf k})=\left(\begin{array}{c c} \epsilon({\bf k})
+Vd_\alpha({\bf k})\sigma^\alpha & \\
& \epsilon({\bf k})-Vd_\alpha({\bf
k})\sigma^\alpha\end{array}\right) \ .
\end{eqnarray}

In other words, the 4-component spin-3/2 system is equivalent to a
decoupled bilayer QAHE system, each with Hamiltonian
(\ref{QAHEHamiltonian}), but with $d_a({\bf k})$ opposite in the
two layers. According to the definition (\ref{conductQAHE}), the
Hall conductivity of the two layers are opposite to each other.
Since $\Gamma^{12}$ is odd under time reversal transformation, the
two layers are time reversal partners, and the total Hamiltonian
$H_{\rm 2d}$ remains time-reversal invariant. The
$\Gamma^{12}$-spin current is given by
\begin{eqnarray}
J_i^{\Gamma}=J_i^+-J_i^-,\qquad i=x,y
\end{eqnarray}
in which $J_i^\pm$ is the current of electrons with
$\Gamma^{12}=\pm 1$. Thus the Hall conductivity of
$J_x^{\Gamma}=\sigma^\Gamma_{xy}E_y$ is quantized as
\begin{eqnarray}
\sigma_{xy}^\Gamma&=&\sigma_{xy}^+-\sigma_{xy}^-
=-\frac{1}{4\pi^2}\int\int_{\rm FBZ} dk_xdk_y {\bf
\hat{d}}\cdot{\bf
\partial_x\hat{d}\times\partial_y\hat{d}}\nonumber\\
&\Rightarrow &\sigma_{xy}^\Gamma=\frac{n}{\pi},\qquad (n\in
\mathbb{Z}) \ .
\end{eqnarray}

\begin{figure}[tbp]
\begin{center}
\includegraphics[width=2.5in] {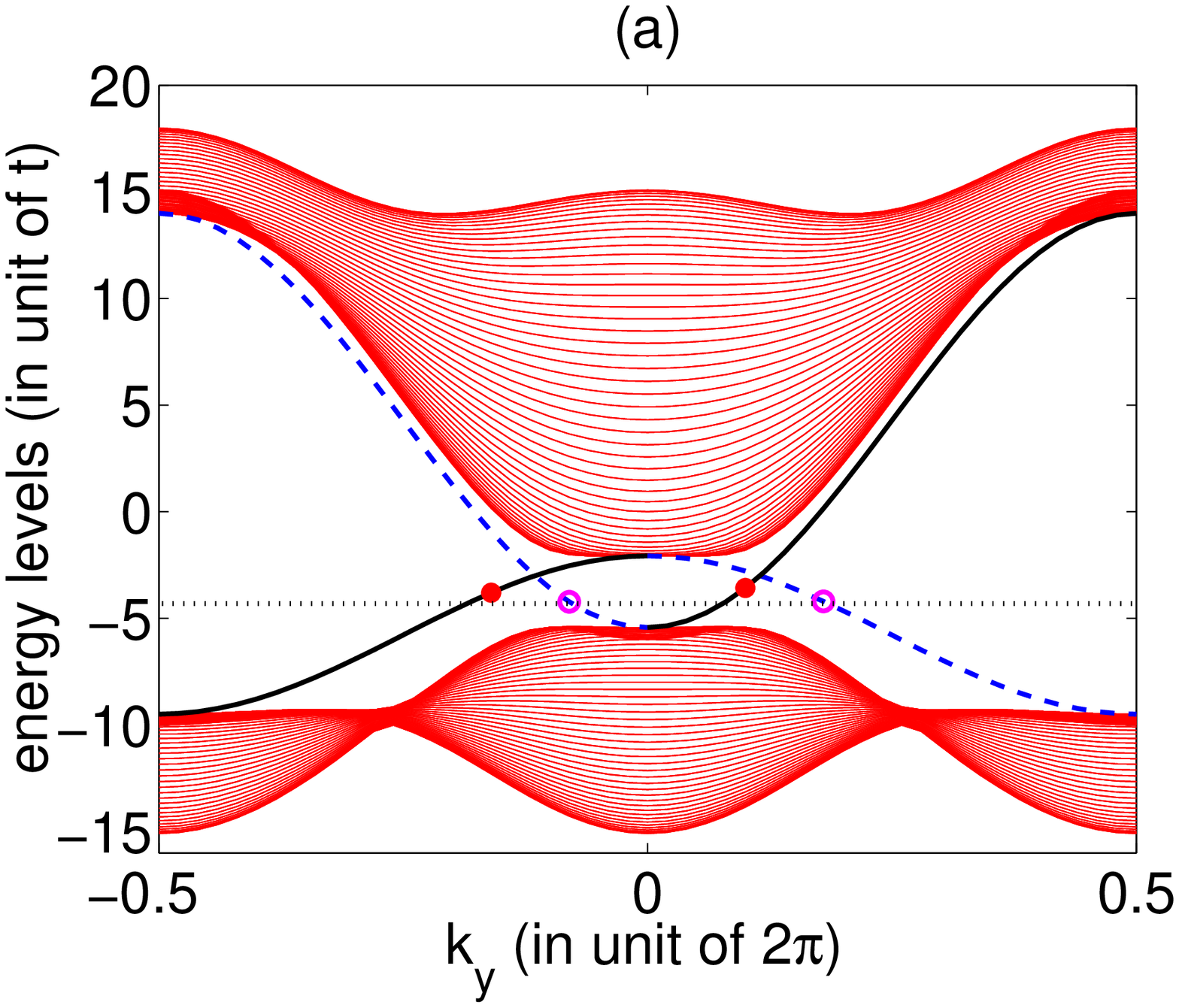}
\end{center}
\begin{center}
\includegraphics[width=2.6in] {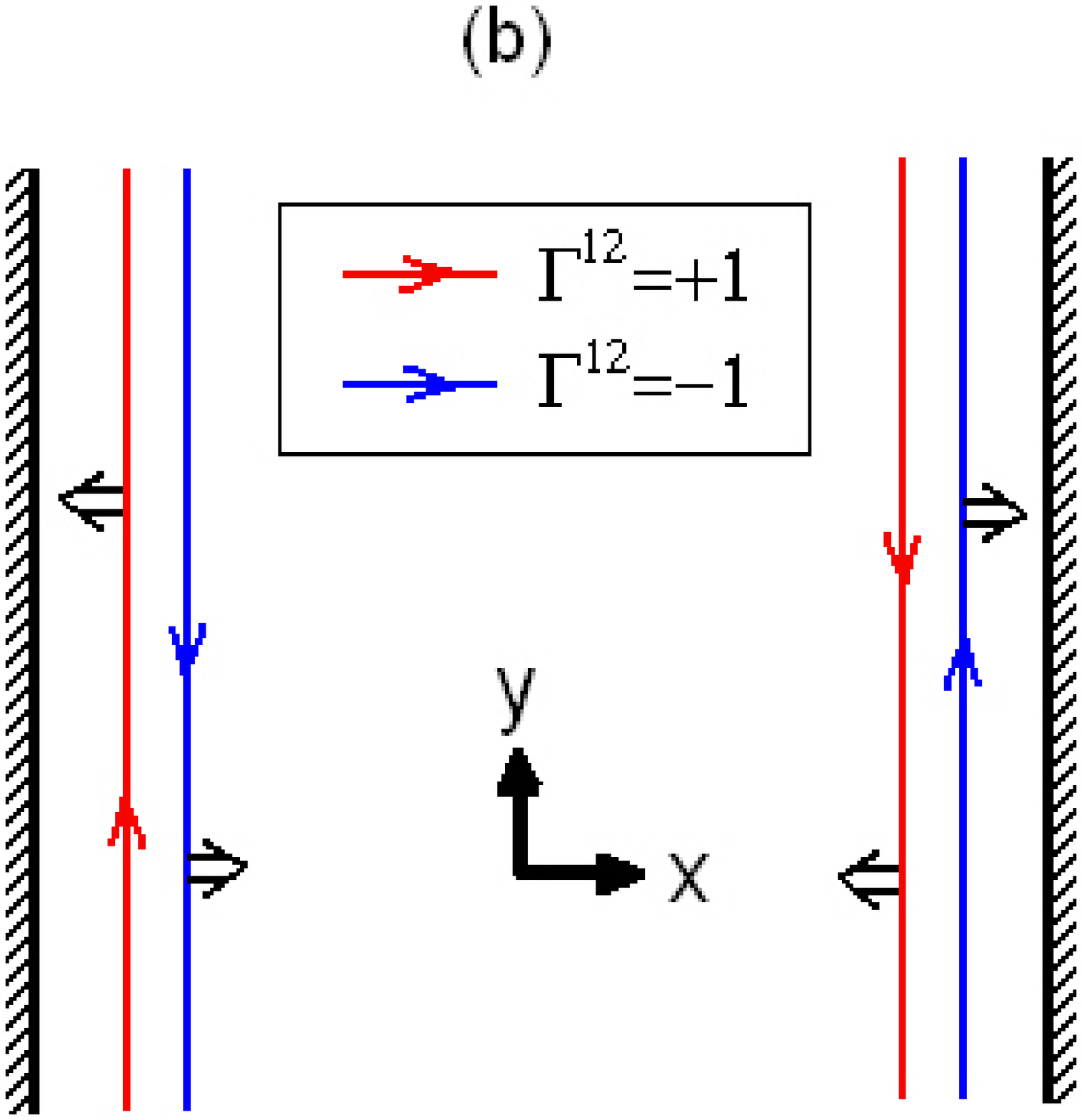}
\end{center}
\caption{(a) The energy spectrum in the case (\ref{choice1}) with
$t/V=4$ and $e_s=0.5$. The isolated solid lines stand for the
(doubly-degenerate) edge states and the dashed line indicates a
typical in-gap fermi energy (with $\mu=-4.2t$). Each crossing of the
fermi energy and the edge state spectrum defines two edge states on
left and right boundary with opposite value of $\Gamma^{12}$. The
solid and hollow circles near a fermi level mark the particle and
hole edge excitations induced by adiabatic flux insertion. (See the
text for details.) (b)Schematic picture of the edge states. Each red
(blue) line stands for an edge state with $\Gamma^{12}=+1$($-1$).
The double arrow shows the pseudo-spin orientation of the current
carried by the corresponding edge state when an electric field is
applied in $y$-direction. For simplicity, only one edge state with
definite $\Gamma^{12}$ eigenvalue on each edge is drawn.}
\label{SHEexample}
\end{figure}

In general, such a quantized Hall conductivity of the conserved
charge $\Gamma^{12}$ leads to a non-vanishing spin Hall effect in
the 2d insulator system (\ref{H2d}), which is consistent with the
three dimensional spin Hall insulator model. What's more, the spin
Hall transport in the two dimensional system can be understood
better by studying the edge states, as in the QHE and QAHE. To see
the picture more clearly, a tight-binding regularization of
$d_a({\bf k})$ is specified as
\begin{eqnarray}
d_3(k)&=&-\sqrt{3}\sin k_x\sin k_y \ , \nonumber\\
d_4(k)&=&\sqrt{3}\left(\cos k_x-\cos k_y\right)\ ,\nonumber\\
d_5(k)&=&2-e_s-\cos k_x-\cos k_y \ ,
\label{choice1}
\end{eqnarray}
which reduces to the continuum form in eqn.(\ref{H2d}) when
$k_x,k_y\rightarrow 0$. Direct calculations show that
\begin{eqnarray}
\sigma_{xy}^{\Gamma}=\left\{\begin{array}{c c}2/\pi, & 0<e_s<4\\0,
& e_s>4\text{ or }e_s<0\end{array}\right. \ , \label{choice1sig}
\end{eqnarray}
thus the topological charge is $2$ when $0<e_s<4$ and $t/V$ is
small. The topological charge in this system is larger than the
previous QAHE example (\ref{QAHEHReal}) by one unit, since the $d$
wave functions here ``winds around" in the momentum space more
than the $p$ wave functions in the previous QAHE example. In this
system there are four edge states on each boundary. For the
$\Gamma^{12}=+1(-1)$ states, the $v_y>0$ state is localized on the
left (right) edge, while the $v_y<0$ state is localized on the
right (left) edge. The energy spectrum and the schematic diagram
of the edge states are shown in Fig. \ref{SHEexample}.

To study the evolution of the edge states in an infinitesimal
electric field, we consider the Laughlin-Halperin gauge
argument\cite{laughlin1981,halperin1982}. The system with the
periodic boundary condition in $y$-direction and open boundary
condition in $x$-direction can be considered as a cylinder. When
the flux $\Phi(t)$ threading the cylinder is adiabatically turned
on from $\Phi(0)=0$ to $\Phi(T)=2\pi$, the electric field in
$y$-direction is given by
\begin{eqnarray}
E_y(t)=-\frac{\partial A_y(t)}{\partial t}=\frac 1 L\frac{\partial
\Phi(t)}{\partial t} \ . \label{Ey}
\end{eqnarray}
For convenience, the direction of the flux $\Phi$ is chosen so
that $E_y>0$ when $\Phi$ increases. The effect of flux threading
can be expressed by replacing $k_y\rightarrow k_y-A_y$ in the
Hamiltonian, or equivalently, as a twisted boundary condition
$\psi(x,y+L)=e^{i\Phi}\psi(x,y)$. In such a picture, each single
particle eigenstate $\phi_{mk_y}(x,y)=u_{mk_y}(x)e^{ik_yy}$ will
be adiabatically transformed into $
\phi_{mk_y}(x,y)(t)=u_{m,k_y-A_y}(x)e^{i(k_y-A_y)y}$. In
particular, when the flux reaches $2\pi$, the adiabatical
evolution will result in
\begin{eqnarray}
\left|m,k_y\right\rangle\rightarrow
\left|m,k_y+\frac{2\pi}L\right\rangle \ .\label{shift}
\end{eqnarray}

When the Fermi level lies in the bulk energy gap, the ground state
of the system is given by $
\left|G\right\rangle=\prod_{E_{mk}\leq\mu}\left|mk\right\rangle$.
When a $2\pi$ flux is threaded through the cylinder, the final
state is obtained by a translation of Fermi sea in momentum space:
$\left|G'\right\rangle=\prod_{E_{mk}\leq\mu}\left|m,k+2\pi/L\right\rangle$
Since the bulk part of $\left|G\right\rangle$ is a product of all
$k$'s, it does not change under such a translation, which implies
that the only difference between $\left|G\right\rangle$ and
$\left|G'\right\rangle$ occurs to the edge states near the Fermi
level. As shown in Fig. \ref{SHEexample} (a) by solid and hollow
circles near the fermi level, each edge state on the Fermi surface
with velocity $v_y>0$ will move out of the fermi sea and becomes a
particle excitation since $\delta E\simeq v_y\delta k=2\pi
v_y/L>0$, while each one with $v_y<0$ will move into the fermi sea
and leads to a hole excitation. Consequently, the final state
$\left|G'\right\rangle$ can be expressed as a particle-hole
excitation state as
\begin{eqnarray}
\left|G'\right\rangle=\prod_{i=1}^nc_{iL+}^\dagger c_{iR-}^\dagger
c_{iL-}c_{iR+}\left|G\right\rangle \ , \label{Gprim}
\end{eqnarray}
in which the label $\pm$ stands for the eigenvalue of
$\Gamma^{12}$ carried by the edge state and $L,R$ refers to the
edge states on the left and right edge, respectively. $n$ is the
bulk topological number. In obtaining eqn. (\ref{Gprim}), we have
used the chirality of the edge states, i.e.
$v_{L+}>0,v_{L-}<0,v_{R+}<0,v_{R-}>0$. (A similar analysis of the
usual IQHE case can be found in Ref. \cite{hatsugai1993}.)

From Eq. (\ref{Gprim}) it's clear that the net effect of
adiabatically turning on a $2\pi$ flux is to transfer edge states
with $\Gamma^{12}=1$ from the right edge to the left one and to
transfer edge states with $\Gamma^{12}=-1$ in the opposite way. This
leads to an accumulation of $\Gamma^{12}$-``spin" on the boundary.
Since $\Gamma^{12}$ is related to $S^z$ by $S^z=-\frac 1
2\Gamma^{12}-\Gamma^{34}$ as shown in Ref.\cite{murakami2004A}, such
an accumulation of $\Gamma^{12}$ in general leads to a non-vanishing
spin $S^z$ density on the boundary. On the other hand, such an
accumulation can also be considered as a consequence of the spin
Hall current $j_x$ induced by the electric field $E_y$ in Eq.
(\ref{Ey}), which implies that the {\em physically observed} spin
Hall conductivity is proportional to the amplitude of spin
accumulation after $2\pi$-flux threading. Since $\left\langle
S^z\right\rangle=-\frac 1 2\left\langle
\Gamma^{12}\right\rangle-\left\langle \Gamma^{34}\right\rangle$, the
corresponding spin Hall conductivity also consists of two parts,
where the conserved part $\sigma_{xy}^{\rm
(c)}=\frac12\sigma^\Gamma_{xy}$ corresponds to a transport of
$\Gamma^{12}$ spin carried by the motion of edge states; while the
non-conserved part $\sigma_{xy}^{\rm (nc)}$ is just a precession
effect due to the non-conserved nature of spin as represented by
$\left\langle \Gamma^{34}\right\rangle$ of each edge state.
Consequently, it is only $\sigma_{xy}^{\rm (c)}$ that counts true
transport of quantum states in the system and is protected by the
bulk topological structure. These considerations give the physical
justification of the conserved spin current operator defined in Ref.
\cite{murakami2004A}.

Finally, we consider the effect of breaking the z-axis mirror
symmetry, which can be induced by adding an asymmetric potential
$U(z)$, such that $U(z)\neq U(-z)$, in the Hamiltonian
(\ref{LuttingerGen}). Consequently, the average $c\equiv\avg{k_z}$
in the lowest 2-d subband becomes finite. And thus an extra term,
\begin{eqnarray}
H_{\rm a}&=&V\left(d_1({\bf k})\Gamma^1+d_2({\bf
k})\Gamma^2\right)\nonumber\\
&=&-\sqrt{3}cV\left(k_y\Gamma^1+k_x\Gamma^2\right)\ ,
\label{assymetric}
\end{eqnarray}
should be added to the two-dimensional Hamiltonian (\ref{H2d}).
Since $\left\{\Gamma^{12},H_{\rm a}\right\}=0$, such a term will
lead to a flip between the states with opposite $\Gamma^{12}$
pseudospin. Especially, a mixing between the left and right moving
edge states will be induced, and thus a gap $E_{\rm edge}\propto
cV$ is open on each edge. Consequently, the system becomes a fully
gapped insulator when the chemical potential $\mu$ lies within the
edge gap; however the gapless edge excitations still exist if
$\mu$ is not in the edge gap but remains within the bulk gap. In
this case, the spin Hall effect carried by the edge states can
still survive, but not as robust as in the fully symmetric case,
since it is not completely topology-protected.

Spin currents in this spin Hall insulator model has also been
discussed in Ref.\cite{Onoda2005}.

\begin{figure}[tbp]
\begin{center}
\includegraphics[width=3in] {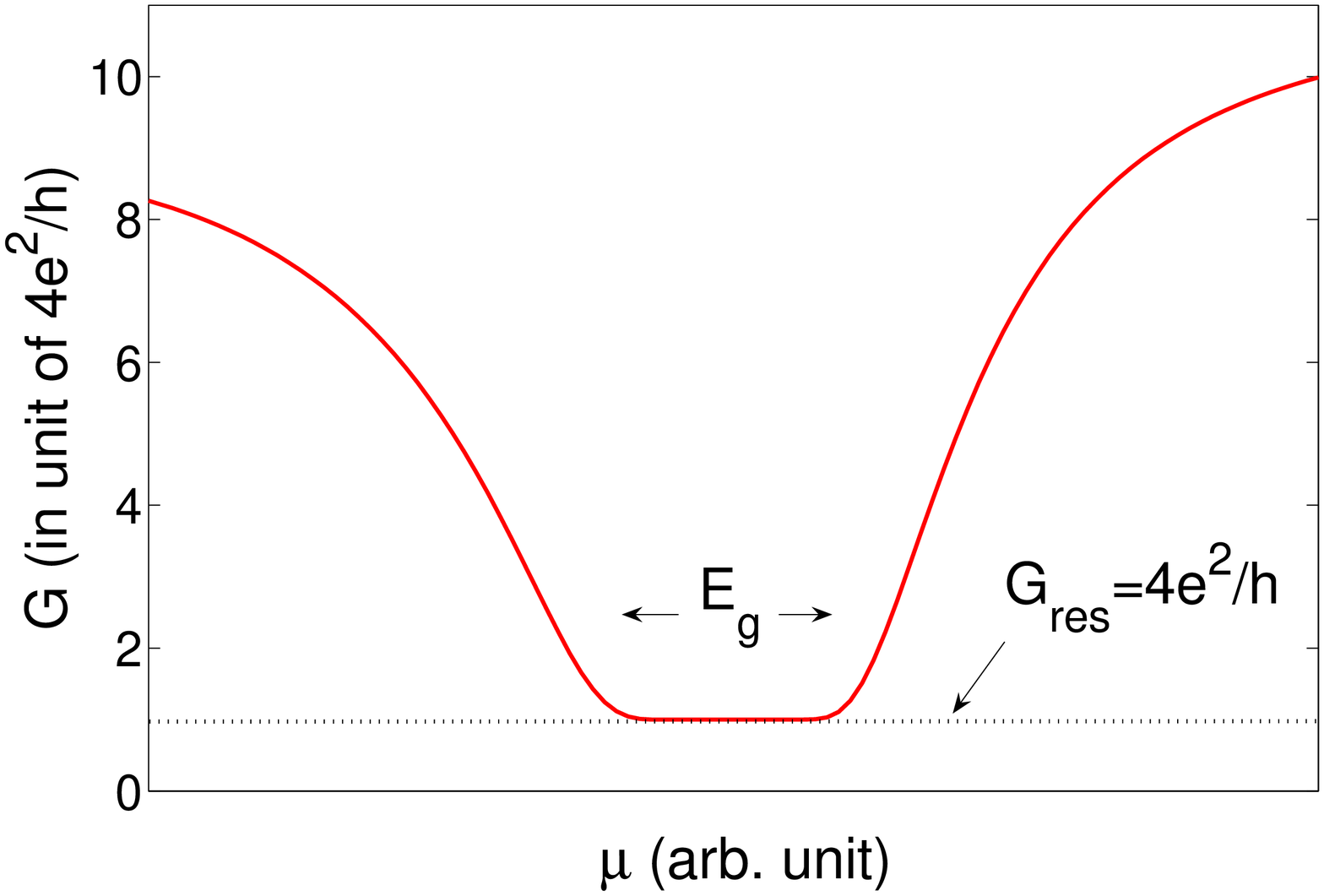}
\end{center}
\caption{The schematic curve of conductance $G$ versus fermi level
$\mu$ for a ballistic quantum well in the quantum spin Hall regime.
The plateau with $G_{\rm res}=\frac{4e^2}{h}$ shows the residual
conductance contributed by 4 edge states on two edges. }
\label{conductance}
\end{figure}

\section{Conclusions and Discussions}

In conclusion, we have proposed a topological mechanism for the
quantum spin Hall effect, which are realized in a class of
two-dimensional spin Hall insulators with mirror symmetry.
Mathematically, such QSHE is the cousin of the quantum Hall effect
in band insulators without a magnetic field, the so-called quantum
anomalous Hall effect (QAHE). By carefully studying the
Laughlin-Halperin gauge argument, insights are gained for the
physical mechanism of the quantum spin Hall transport carried by
the edge states. It also provides an understanding of the physical
meaning of spin transport in the absence of spin conservation.

The QSHE models discussed in this paper can be experimentally
realized in two classes of 2D semiconductors. One class is the
(distorted) zero-gap semiconductors such as HgTe, HgSe, $\beta$-HgS
and $\alpha$-Sn. The other class is the narrow-gap semiconductors
such as PbTe, PbSe and PbS\cite{murakami2004B}. As proposed above,
topological quantization of the spin Hall effect shows up in the
cases with mirror symmetry with respect to the z-axis, realizable
when the 2D material is trapped in a symmetric quantum well. Once
the quantum spin Hall effect is realized in the ground state, it is
protected against thermal fluctuations by the bulk energy gap $E_g$.
At temperature $T\ll E_g/k_B$, the quantum spin Hall effect is
expected to be observable. Take ${\rm HgTe/Hg_{1-x}Cd_xTe}$ (001)
quantum well for an example. As calculated in Ref\cite{novik2005},
the gap $E_g$ between LH and HH bands is of order $10{\rm meV}$ for
$x=0.7$, which means the quantum spin Hall effect can be observed in
a wide temperature range $T\ll 100K$. Once the spin Hall effect in
the pure system is established, it is not significantly dependent on
the mobility of the material.

However, the magnitude of edge spin accumulation in a steady state
is in general dependent on the spin relaxation mechanism and the
disorder in the system. To avoid ambiguity in estimating the edge
spin accumulation, here we propose a more definite experimental
prediction of the QSHE. As is shown above, there are two pairs of
gapless edge states on each boundary, which are chiral in spin
transport but {\em nonchiral} in the charge channel. In other
words, each pair of edge states (with opposite $\Gamma^{12}$ on
the same boundary) is equivalent to a {\em spinless} Luttinger
liquid in two-terminal measurements of charge conductance.
Consequently, there are in total 4 pairs of edge states in the
system, which  will make a quantized
contribution\cite{maslov1995},
\begin{eqnarray}
G=\frac{4e^2}{h} \ ,
\end{eqnarray}
to the longitudinal charge conductance when the system is in a
ballistic regime. The schematic curve of longitudinal conductance
versus the Fermi level is shown in Fig. \ref{conductance}. Compared
to the vanishing conductance in a trivial insulator, such a residual
conductance provides a simple probe of the topologically non-trivial
edge states in the QSHE.

We would like to acknowledge helpful discussions with Andrei
Bernevig, C. L. Kane, A. H. MacDonald, Z. Y. Weng and Cenke Xu .
This work is supported by the NSFC under grant numbers 10374058
and 90403016, the US NSF under grant numbers DMR-0342832 and
PHY-0457018 and the US Department of Energy, Office of Basic
Energy Sciences under contract DE-AC03-76SF00515.

\appendix
\section{Kubo formula calculation of $\sigma_{xy}$}

Here we propose a derivation of eqn. (\ref{conductQAHE}) from the
Kubo formula. The single particle Green function corresponding to
the Hamiltonian (\ref{QAHEHamiltonian}) can be written as
\begin{eqnarray}
G\left({\bf k},i\omega_n\right)&=&\left(i\omega_n-H({\bf
k})\right)^{-1}\nonumber\\
&=&\frac{P_+}{i\omega_n-E_+({\bf
k})}+\frac{P_-}{i\omega_n-E_-({\bf k})}\ ,
\label{Greenfunction}\\
\text{with } P_\pm&=&\frac 1 2\left(1\pm \hat{d}_\alpha({\bf
k})\sigma^\alpha\right) \ .\nonumber
\end{eqnarray}
Then the charge Hall conductivity can be calculated using the Kubo
formula (\ref{kubo}):
\begin{widetext}
\begin{eqnarray}
Q_{xy}(i\nu_m)&=&\frac{1}{\Omega\beta}\sum_{{\bf k},n}{\rm
tr}\left(J_{x}({\bf k})G({\bf
k},i(\omega_n+\nu_m))J_y({\bf k})G({\bf k},i\omega_n)\right)\nonumber\\
&=&\frac{1}{\Omega\beta}\sum_{s,t=\pm}\sum_{{\bf k},n}\frac{{\rm
tr}\left(J_{x}({\bf k})P_s({\bf k})J_y({\bf k})P_t({\bf
k})\right)}{\left(i\left(\omega_n+\nu_m\right)-E_s({\bf
k})\right)\left(i\omega_n-E_t({\bf
k})\right)}\nonumber\\
&=&\frac{1}{\Omega}\sum_{s,t=\pm}\sum_{{\bf k}}\frac{{\rm
tr}\left(J_{x}({\bf k})P_s({\bf k})J_y({\bf k})P_t({\bf
k})\right)}{i\nu_m-E_s({\bf k})+E_t({\bf k})}\left(n_t({\bf
k})-n_s({\bf k})\right) \ , \nonumber\\
\Rightarrow \sigma_{xy}&=&\lim_{\omega\rightarrow 0}\frac i\omega
Q_{xy}(\omega+i\delta)=-\frac{i}{\Omega}\sum_{s,t=\pm}\sum_{{\bf
k}}\frac{{\rm tr}\left(J_{x}({\bf k})P_s({\bf k})J_y({\bf
k})P_t({\bf k})\right)}{\left(E_t({\bf k})-E_s({\bf
k})\right)^2}\left(n_t({\bf k})-n_s({\bf k})\right)\nonumber\\
&=&-\frac{i}{\Omega}\sum_{{\bf k}}\frac{n_-({\bf k})-n_+({\bf
k})}{4V^2d({\bf k})^2}\left[{\rm tr}\left(J_{x}({\bf k})P_+({\bf
k})J_y({\bf k})P_-({\bf k})\right)-h.c.\right] \ .
\label{kubo2}
\end{eqnarray}
Here $E_\pm ({\bf k})=\epsilon({\bf k})\pm Vd({\bf k})$ is
introduced in the last line. The trace in eqn. (\ref{kubo2}) can
be worked out by substituting $J_{x,y}({\bf k})$ and $P_\pm({\bf
k})$ with their definition (\ref{currentoperator}) and
(\ref{Greenfunction}), respectively:
\begin{eqnarray}
\sigma_{xy}&=&\frac i {4\Omega}\sum_{k}{\rm
Tr}\left\{\left(\frac{\partial \epsilon(k)}{\partial
k_x}+V\frac{\partial d_\alpha(k)}{\partial
k_x}\sigma^\alpha\right)\left(1-\hat{d}_\alpha\sigma^\alpha\right)
\left(\frac{\partial \epsilon(k)}{\partial k_y}+V\frac{\partial
d_\alpha(k)}{\partial k_y}\sigma^\alpha\right)
\left(1+\hat{d}_\alpha\sigma^\alpha\right)-h.c.\right\}
\frac{n_-({\bf k})-n_+({\bf k})}{4V^2d^2}\nonumber\\
&=&-\frac 1{2\Omega}\sum_{k}\left\{\frac{\partial
\hat{d}_\alpha(k)}{\partial k_x}\frac{\partial
\hat{d}_\beta(k)}{\partial
k_y}\hat{d}_\gamma\epsilon^{\alpha\beta\gamma}\right\}\left(n_-({\bf
k})-n_+({\bf k})\right) \ .
\label{conductcalculation2}
\end{eqnarray}
\end{widetext}
For the last step, it should be noticed that only the
three-$\sigma^\alpha$ terms in the expansion make non-vanish
contribution to the trace. Thus the final result is independent of
coupling $V$, as expected from topological considerations. This
finishes our derivation of the formula (\ref{conductcalculation})
for the spin Hall conductance.

\bibliography{QSHE,QHE}

\end{document}